# Interaction of edge exciton polaritons with engineered defects in the van der Waals material Bi$_2$Se$_3$

Robin Lingstädt*,1, Nahid Talebi*,1,4, Mario Hentschel2, Soudabeh Mashhadi1, Bruno Gompf3, Marko Burghard1, Harald Giessen2 and Peter A. van Aken1

1 Max Planck Institute for Solid State Research, Stuttgart, Germany
2 4th Physics Institute, University Stuttgart, Germany
3 1st Physics Institute, University Stuttgart, Germany
4 Institute for Experimental and Applied Physics, Christian Albrechts University, Kiel, Germany
* corresponding authors and have equally contributed to this work

Hyperbolic materials exhibit unique properties that enable a variety of intriguing applications in nanophotonics. The topological insulator Bi$_2$Se$_3$ represents a natural hyperbolic optical medium, both in the THz and visible range. Here, using cathodoluminescence spectroscopy and electron energy-loss spectroscopy, we demonstrate that Bi$_2$Se$_3$, in addition to being a hyperbolic material, supports room-temperature exciton polaritons. Moreover, we explore the behavior of hyperbolic edge exciton polaritons in Bi$_2$Se$_3$. Edge polaritons are hybrid modes that result from the coupling of the polaritons bound to the upper and lower edges of Bi$_2$Se$_3$ nanoplatelets.

In particular, we use electron energy-loss spectroscopy to compare Fabry-Pérot-like resonances emerging in edge polariton propagation along pristine and artificially structured edges of the nanoplatelets. The experimentally observed scattering of edge polaritons by defect structures was found to be in good agreement with finite-difference time-domain simulations. Moreover, we experimentally proved coupling of localized polaritons in identical open and closed circular nanocavities to the propagating edge polaritons. Our findings are testimony to the extraordinary capability of the hyperbolic polariton propagation to cope with the presence of defects. This provides an excellent basis for applications such as nanooptical circuitry, cloaking at the nanometer scale, as well as nanoscopic quantum technology on the nanoscale.

In the past few years, anisotropic media with a hyperbolic dispersion have attracted significant attention due to their unique electromagnetic and optical properties[1]. A material is called hyperbolic, when its isofrequency surface given by

$$\frac{k_{\parallel}^2}{\varepsilon_{\perp}} + \frac{k_{\perp}^2}{\varepsilon_{\parallel}} = \left(\frac{\omega}{c}\right)^2 \qquad (1)$$

forms a hyperboloid. Here, $c$ is the speed of light and $k_{\parallel}$ and $k_{\perp}$ are the in-plane (*x-y*-plane) and out-of-plane (*z*-direction) components of the wave vector, respectively, describing the plane-wave propagation in the material at frequency $\omega$. This hyperbolic condition is fulfilled when at least one principle component of the dielectric tensor $\varepsilon$ is negative[2]. A material with a negative real part of the dielectric function in the isotropic plane parallel to the surface ($\varepsilon_{\parallel}$) is referred to as hyperbolic type-2 (HB$_2$), whereas a material displaying a negative value in the out-of-plane direction ($\varepsilon_{\perp}$) is of hyperbolic type-1 (HB$_1$) character. In contrast to closed isofrequency sheets, in hyperbolic materials propagation can occur at arbitrarily large wave vectors, resulting in a number of peculiar nanophotonic properties and applications such as negative refraction[3] or subdiffraction super-resolution imaging[4-7] right at the transition to the hyperbolic dispersion[2]. The main criterion for hyperbolicity is the motion of free electrons being constrained in one or two spatial dimensions, which can be experimentally realized by the construction of artificial metamaterials such as layered metal-dielectric structures[8-11] or based on silicon carbide particles[12]. Besides such artificial engineering, and even more intriguing, natural hyperbolic behavior has also been predicted and experimentally demonstrated, for example in graphite for ultraviolet frequencies[13], in layered van der Waals materials in the visible range[14], and hexagonal boron nitride (h-BN) in the terahertz regime[15-18]. Tetradymites like $Bi_2Se_3$ structures are another class of hyperbolic materials in the visible range, as confirmed by spectroscopic ellipsometry[19,20]. Moreover, $Bi_2Se_3$ can support the propagation of long-range hyperbolic edge polaritons[21], which are excited at the edges of nanoplatelets (Fig. 1a). Being topologically different from surfaces, edges support polaritons with distinct features, such as having ultrahigh confined mode volumes and hybrid optical responses[22]. It should be mentioned though, that in contrast to directional zigzag-like propagation and scattering of surface phonon polaritons along the side surfaces of h-BN nanoparticles, the investigated edge polaritons here are hybrid modes that are tightly bound to the edges of the material, rather than 2D surfaces.

We demonstrate for the first time – using cathodoluminescence spectroscopy (see Supplementary Information Fig. 1) – that $Bi_2Se_3$ supports room-temperature excitons, with binding energies higher than $E = k_B T$, where $k_B$ is the Boltzmann constant and $T$ is the temperature. Although polaritons in $Bi_2Se_3$ and $Bi_2Te_3$ have been demonstrated before[21,23,24], the origin of the observed polaritonic behavior had not been revealed. Similar direct and indirect band gap excitonic excitations with strong oscillator strengths up to the room temperature exist in other either engineered heterostructures[25,26] or natural layered materials with van der Waals bindings[27-31]. Combination of the hyperbolic responses and exciton excitations thus leads to the propagation of hyperbolic edge exciton polaritons (HEEPs). In addition, we investigate the interaction of HEEPs with engineered defects, as means to steer edge polaritons. Of particular relevance is the development of reliable means to modulate the reflection and transmission behavior of HEEPs at corners of nanostructures. Here, we manipulated side planes of 50 nm-thin $Bi_2Se_3$ nanoplatelets with inhibited grooves or specifically shaped nanocavities using focused ion beam milling, with the aim of exploring the transmission and scattering behavior of HEEPs and their interaction with localized polaritons. To this end, we use electron energy-loss spectroscopy (EELS)[32,33] in a transmission electron microscope (TEM) specifically adapted for investigations in the low-loss energy range, and compare the experimental data with results of finite-difference time-domain (FDTD)[34] simulations. We furthermore demonstrate that high-energy HEEPs, owing to their coupling to radiation modes via various mechanisms of energy transfer, are rather robust against the engineered defects.

**Results**

We first investigate $Bi_2Se_3$ nanoplatelets using cathodoluminescence (CL) spectroscopy. Electrons traversing semiconducting materials undergo a series of inelastic events. Particularly, a number of electron-hole pairs is generated per electron excitation, which strongly depends on the energy of the incident electron and the band gap of the material. Therefore, CL is a deterministic tool for detecting exciton-mediated radiation of the material. Acquired CL spectra from the material (see Supplementary Information Fig. 1), particularly show two exciton peaks at the energies of $E = 1.5$ eV and $E = 1.95$ eV. These peaks are associated with the exciton transitions around the Q and F points of the Brillouin zone[35]. Exciton excitation in between the quintuple layers (located at the Γ point of the Brillouin zone) happens at lower energies below 0.5 eV, and their energies are affected by the spin-orbit interactions, similar to the exciton peaks in $MoS_2$[36]. However, this energy range is out of reach of the used CL spectrometer. The combination of excitonic excitations together with the hyperbolic nature of this material, thus leads to the emergence of hyperbolic exciton polaritons and in particular HEEPs. In the followings, the properties of HEEPs and their interaction with anomalies will be discussed.

Thin nanoplatelets of $Bi_2Se_3$ were investigated further experimentally by means of analytical TEM. For that purpose, the area of interest is either irradiated with a parallel beam of fast electrons or scanned by a focused electron probe. In both cases, the electrons lose energy due to the inelastic interaction with the specimen, and their corresponding energy-losses are analyzed with an energy-dispersive detection system (Fig. 1 a). Electron beams interacting with nanoplatelets in an aloof trajectory, i.e., traversing the mode volume surrounding the side planes without passing through the material, can launch both surface waves and HEEPs. As will be discussed here, we observe a stronger coupling of the electron beams to HEEPs rather than surface waves. HEEPs in nanoplatelets propagate along the edges with their mode volume being sharply confined to the edges (Fig. 1 b).

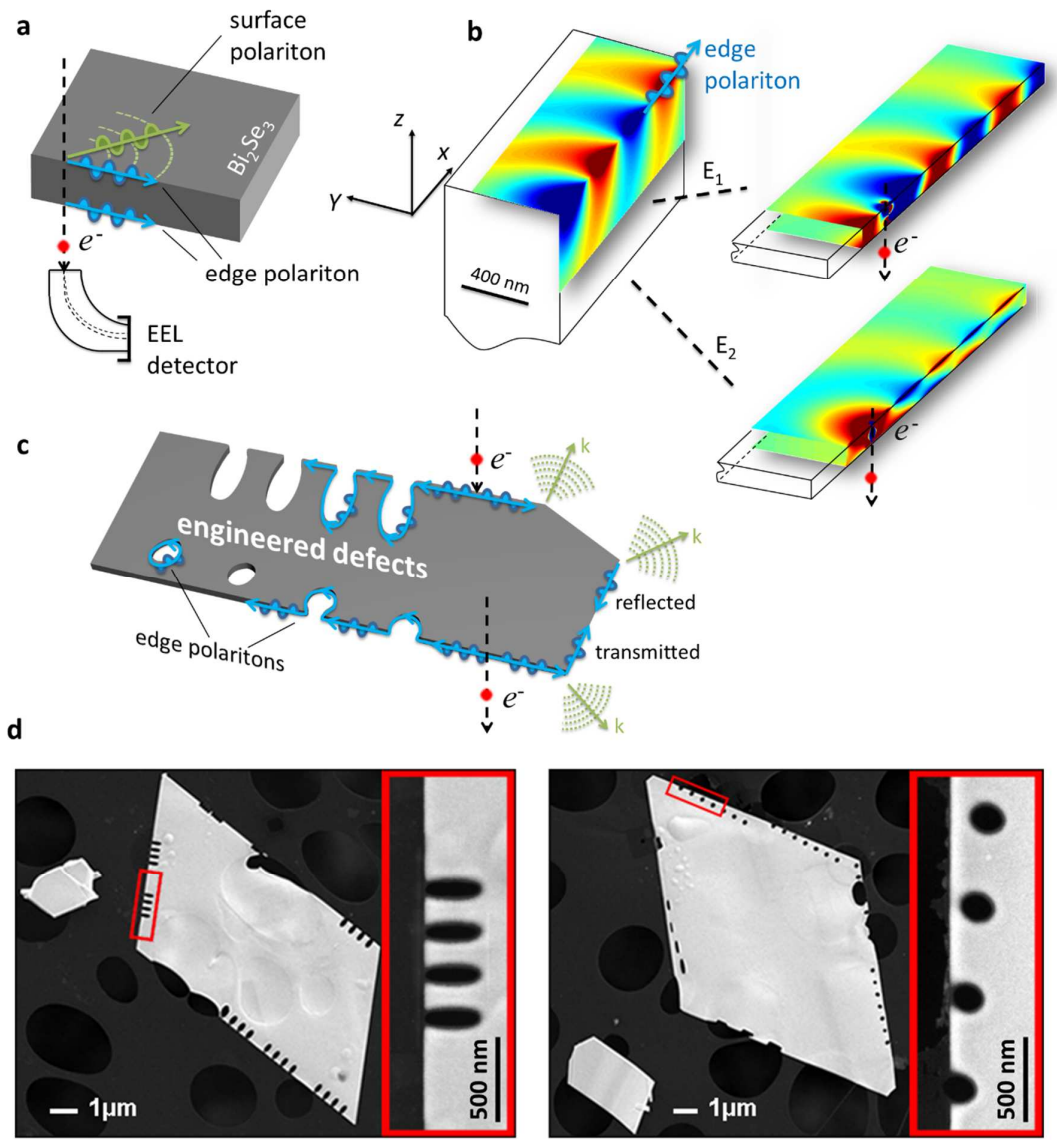

**Fig. 1 Excitation of polaritons in $Bi_2Se_3$ by electron beams.** (a) During the inelastic interaction of fast electrons with the specimen, both surface polaritons and edge polaritons are excited, and consequently the electrons experience energy-losses. The electron energy-loss spectrum is then acquired by the energy-dispersive detection system. (b) *z*-component of the electric field associated with edge polatitons propagating along the edges of a $Bi_2Se_3$ large cube, at the energy of 4 eV. The edge modes are hybridized in thin films to form symmetric and antisymmetric modes. Here, $E_1$ = 4.3 eV, and $E_2$ = 3.8 eV, and the thickness of the film is 60 nm. (c) The interaction of hyperbolic edge polaritons at corners and with precisely engineered scatterers of different topology causes reflection, transmission, and radiation of partial waves. (d) SEM images of the investigated structures.

The HEEPs at the upper and lower edge of the nanoplatelets hybridize into a symmetric and an anti-symmteric mode, very similar to surface polaritons in thin films.

Launched optical modes that propagate along the edges of a nanoplatelet are partially reflected at corners, partially guided around them to the adjacent side planes, or couple to far-field radiation (Fig. 1 b). The realization of nanocircuitry devices for applications such as polaritonic cloaking[37] in the general field of transformational flat optics[38,39] requires tunable steering of the propagating modes to desired locations under the possible influence of interactions with local topological anomalies.

Here, we aim at understanding the interaction of HEEPs with such anomalies. For this purpose, different defect structures were created at clean edges of $Bi_2Se_3$ nanoplatelets using focused ion beam milling. In the present paper, two examples are discussed, which are sketched in Fig. 1 c. The real structures are located on two different nanoplatelets, as displayed in scanning electron microscope (SEM) images in Fig. 1 d. Binary images have been extracted as a topological basis for FDTD simulations.

First, we provide a detailed mode analysis for the HEEPs. Due to the anisotropic dielectric properties of the material (Fig. 2 a), it becomes hyperbolic within distinct energy ranges between 1.06 eV and 1.74 eV and above 1.9 eV with type-1 and type-2 character, respectively. The aforementioned partial reflection of propagating modes results in the formation of standing wave patterns with a finite standing wave ratio. This modulation of the photonic local density of states is visible in the EELS signal[40,41]. Fabry-Pérot-like resonances of several orders were observed along edges at distinct energy values, depending on their length (see Supplementary Information Fig. 2). From the distance between the maxima, the wavelengths and corresponding propagation constants of the excited modes were extracted, which show very good agreement with the computed dispersion (Fig. 2 b and Supplementary Information Fig. 2 c). At energies below 1.06 eV, channel modes are excited that have field profiles confined to the upper and lower surfaces. At energies starting from 1.06 eV, when the material becomes hyperbolic, HEEPs are excited with longer propagation lengths (Fig. 2 b, inset) and therefore reduced attenuation constants, particularly at energies above 3.0 eV.

Two coexisting HEEP modes with symmetric and antisymmetric field distributions are supported (Fig. 2 c). Both modes are hybrid in nature, meaning that they cannot be described by either transverse magnetic or transverse electric mode profiles, but rather a superposition of both solutions are required. Intriguingly, they can be described by HE mode profiles (see supplementary note 2). $HE_1$ and $HE_2$ modes are associated with antisymmetric and symmetric HEEPs respectively, where by symmetry, we refer to the spatial distribution of the induced charges.

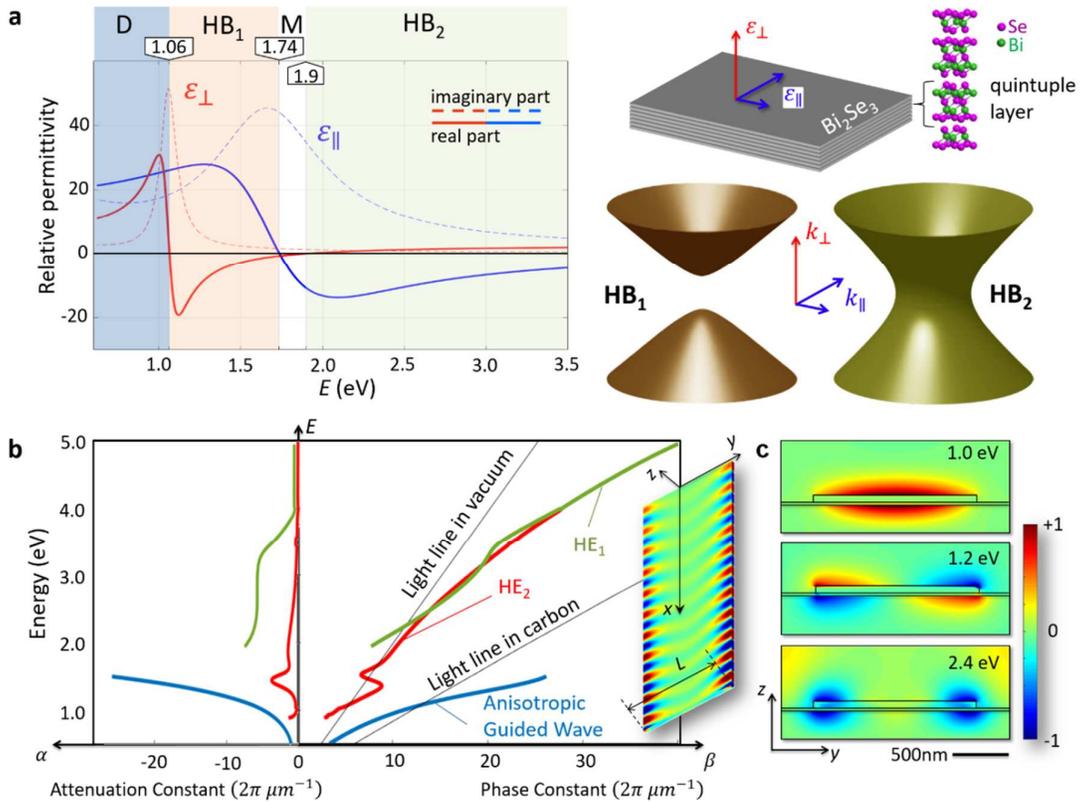

**Fig. 2 Hyperbolic properties of Bi$_2$Se$_3$ due to high spatial anisotropy.** (a) Left: Relative permittivity of bulk Bi$_2$Se$_3$, acquired by ellipsometry measurements. Anisotropic guided waves (D), hyperbolic type-1 (HB$_1$) and hyperbolic type-2 (HB$_2$) waves are supported at the energies highlighted by blue, orange, and green colors, respectively. Right: In-plane and out-of-plane directions within the layered van der Waals crystal and isofrequency sheets related to hyperbolic type-1 and type-2 behavior. (b) Dispersion diagram of the different optical modes, supported by a Bi$_2$Se$_3$ nanoplatelet in the form of a channel waveguide (only lower order modes are shown; for a full diagram including experimental data see Supplementary Information Fig. 2). HE$_1$ and HE$_2$ indicate hybrid magnetic-like modes. The inset graphic shows the simulated spatial distribution of the *z*-component of the electric field associated with the HE$_1$ mode, that propagates in *x*-direction along an infinitely long waveguide. (c) The *z*-component of the electric field at the cross sections associated with the anisotropic guided wave, hyperbolic type-1 HE$_2$ and hyperbolic type-2 HE$_1$ channel modes, are shown at the depicted energies from top to bottom, respectively.

The structure with round-shaped grooves (Fig. 3 a) was experimentally analyzed by energy-filtered transmission electron microscopy (EFTEM), where a series of energy-filtered images within a certain energy-loss range is recorded at intervals of 0.2 eV. These images form a three-dimensional data cube with the *x* and *y* coordinates as spatial axes and the *z* coordinate as the energy-loss axis. The intensity variation in each energy slice visualizes the spatially resolved relative energy-loss probability, which is proportional to the induced electric field projected along the electron trajectory (Fig. 3 b)[42].

Indeed, using EFTEM mode for spectral imaging rather than scanning the structure with a focused electron beam yields better spatial sampling over this large investigated area and reduces the risk of contamination and beam damage. While there is no spatial intensity modulation present at 1 eV (data not shown here), a high intensity is observed along the structured side planes with maxima inside the grooves and minima at the corners at 3 eV (Fig. 3 c and d). They are marked with big and small blue arrows in the EFTEM image (Fig. 3 b, left panel). These intensity modulations are caused by the interaction of $HE_1$ and $HE_2$ modes by the discontinuities like sharp corners and partial reflection from them. Lack of intensity modulations at 1 eV is understood from the fact that at this energy, the dispersion of the HEEPs is located inside the light cone (Fig. 2 b); therefore, electron-induced excitations strongly couple to the radiation. The perimeter of each groove structure is approximately 700 nm, that is equal to $2\lambda_{\text{eff}}$ at this energy, where $\lambda_{\text{eff}} = 2\pi/\beta(\omega)$ is the effective HEEP wavelength and $\beta(\omega)$ is the phase constant. Interestingly, despite the relatively large size of the groove, only one single maximum is observed. We associate this behavior to the coexistence of two $HE_1$ and $HE_2$ modes at this energy with slightly different propagation constants, in such a way that interference between these two optical modes leads to a beating frequency observed as a low frequency spatial modulation. At 4 eV, absorption maxima are measured at the narrow bridges between the grooves (big blue arrows in Fig. 3 b, right panel). At this energy, $HE_1$ and $HE_2$ modes become degenerate, and the effective wavelength associated with both modes is approximately 350 nm, that is two times the length of the narrow bridges. At energies higher than 4 eV, the EELS signal for electrons traversing the $Bi_2Se_3$ film becomes more prominent, highlighting the existence of surface plasmons (see supplementary Fig. 5).

Additional measurements were performed on a second finite grating structure, with grooves separated by 300 nm gaps from the edge, (Fig. 4 a). EEL spectra were acquired in scanning TEM mode in an aloof probe position, where the electron probe is positioned in close vicinity to the specimen[43]. Indeed, in a modal description, electron beams can couple both to surface and edge polaritons. In addition, also scattering of surface waves at the edges excites polaritons propagating along the edges. In an aloof experiment, like the EELS scan data in Fig. 4, electrons can only couple to edge polaritons. A spectral line scan along the structured edge was obtained from a spectrum image by averaging over four pixel rows to increase the signal-to-noise ratio. Subtracting a first-order log-polynomial background, fitted to the tail of the monochromated zero-loss peak within Digital Micrograph software, resulted in the EELS intensity distribution of Fig. 4 b.

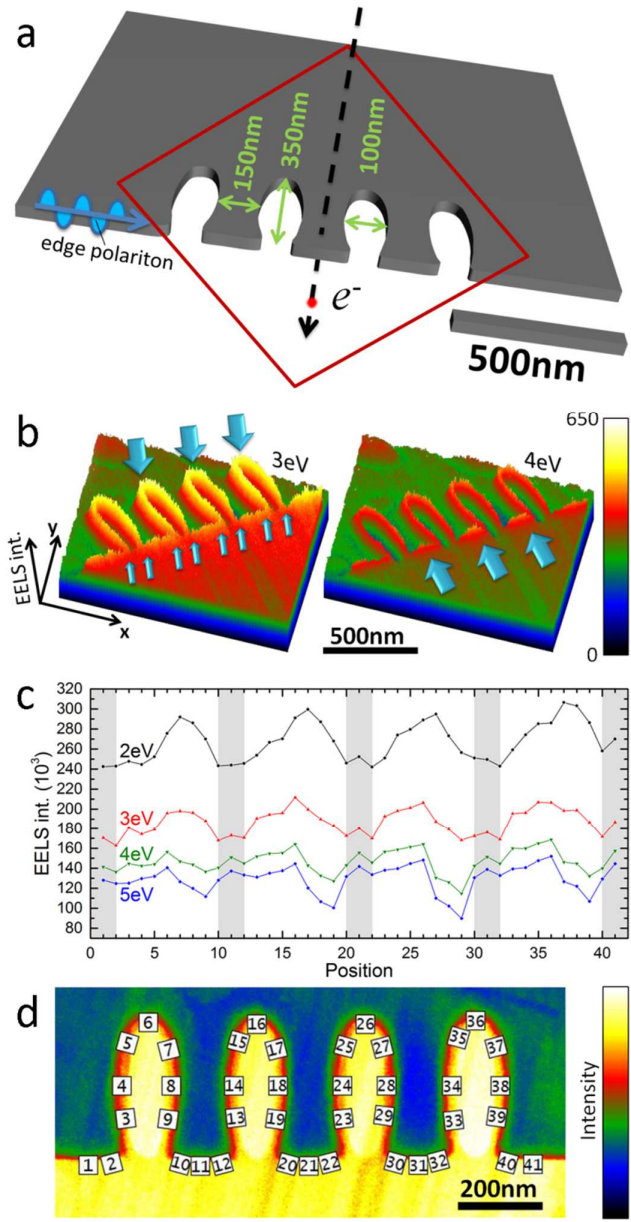

**Fig. 3 Energy-filtered TEM images of the finite grating structure at a Bi$_2$Se$_3$ edge.** (a) Schematic of the finite grating structure, consisting of round-shaped grooves. The investigated area is marked by a red frame. (b) EFTEM images at different energy-losses (3 eV and 4 eV) acquired with an energy window of 0.2 eV width. The intensity variation indicates the relative local absorption probability. Big and small blue arrows point towards intensity maxima and minima, respectively, which are referred to in the main text. (c) Intensity modulation along the structured edge for different energy-loss values as indicated within the figure. The analyzed spectra were extracted from positions marked in (d) with numbers and averaged spatially over an area of 10x10 pixels. Grey shaded areas in (c) emphasize the positions of the bridges between grooves.

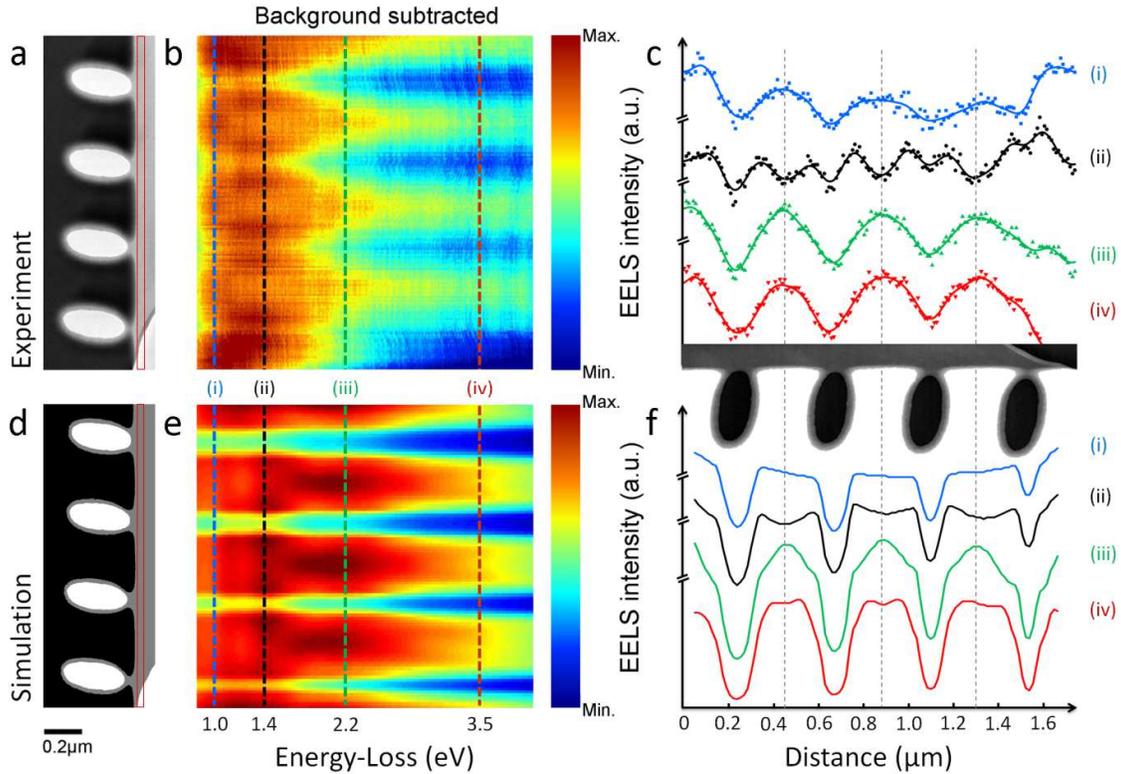

**Fig. 4 EELS line scan along a Bi$_2$Se$_3$ edge with a finite grating structure.** (a) Bright-field TEM image of the nanostructure. It is composed of an edge with elliptical grooves. The form of the boundaries was extracted as a binary image (d) and used for simulations. (b,e) Experimental and simulated energy-distance maps showing absorption probabilities depending on energy-loss value and probe position were extracted from the area marked with a red box in the survey image. Line profiles at selected energies (1.0 eV blue, 1.4 eV black, 2.2 eV green, 3.5 eV red) are shown in (c) and (f). They are shifted vertically for clarity. The experimental data (c) were processed by applying a Fourier transform smoothing algorithm in the plotting software, using a 5-point window (solid lines).

The color-coded absorption probabilities versus the probe position show intensity maxima at the energy-losses of $\Delta E$ = 1 eV and $\Delta E$ = 2.2 eV at the centers and at $\Delta E$ = 1.4 eV at the corners of the bridges, as confirmed by extracted spectra at these positions (see Supplementary Information Fig. 7). The spatial modulation of distinct energy-loss values along the edge was extracted as line profiles (c) that are in good agreement with the simulations (f). Additionally, at energy ranges $E < 1$ eV, and $1.1 \text{ eV} < E < 1.7$ eV, we observe the first-order and second-order Fabry-Pérot-like resonances with one and two intensity maxima along the bridge, respectively (Fig. 4 b and c). At energies $1.7 \text{ eV} < E < 2.2$ eV however, we observe again a first-order mode. This is particularly due to the change in the material dispersion from HB$_1$ to HB$_2$ and also corresponding changes in the phase constant and effective wavelength of the hyperbolic polaritons.

To investigate the propagation around defects in detail, open and closed circular nanocavity structures were created at the edges of nanoplatelets to study the interference of localized polaritons with propagating modes. Holes with a diameter of ca. 200 nm and a side-to-side distance of ca. 350 nm were milled at different distances from the edge, resulting in two open cavities with an opening gap of about 150 nm and 100 nm, and edge distances of ca. 50 nm, 100 nm, and 150 nm. A three-dimensional sketch of the structure is shown in Fig. 5 a. The red frame marks the area containing the first three holes, which has been investigated via scanning TEM-EELS spectrum imaging, where the electron probe is scanned over the specimen, while detecting a complete electron energy-loss spectrum for each pixel. Color-coded energy-filtered images have been extracted from the 3D data cube at selected energy-loss values in the range between 1 eV and 4 eV. As for the finite grating of the grooves discussed before, the EELS signal is mainly confined to the shaped edge of the nanoplatelet and the intensity exhibits maxima within the holes and minima at the corners. Interestingly, this is also the case for the isolated hole, which is located at a distance of 50 nm away from the edge. As discussed before, Fabry-Pérot-like resonances that occur at finite edges are detectable via spatial intensity modulations in the EELS signal. This intensity modulation is observed along the edge between the two open holes with a maximum in the center for 1 eV, two maxima at the corners for 1.5 eV and three maxima (one in the center and two at the corners) for 2 eV. Interestingly, the same modulations, though with less visibility, are detected along the edge between the second and the third hole, which clearly reveals an influence of the latter isolated resonator on the propagating edge polaritons. EELS spectra that have been extracted from several edge positions clearly reveal the resonance character of the observed intensity modulations and are shown in the Supplementary Information Fig. 8.

Simulated spectral line scans confirm these experimental findings (Fig. 5 b-e). Nevertheless, the contrast of the EELS signal along the edge relative to the EELS signal from the open cavities is more pronounced in the FDTD simulations. At higher energies, this contrast is significantly altered, which is due to an enhanced coupling between the edge polaritons and localized SPs within the nanocavities, as will be discussed later. Moreover, at energies above 4 eV, there is no significant intensity modulation observable along the edges at the location of the $3^{rd}$ and $4^{th}$ isolated nanocavity. Nevertheless, even at energies below 4 eV, the influence of the third hole on the intensity modulations is more pronouncedly observed in the experimental results when compared to the simulations. This discrepancy can have several reasons: (i) Due to the weak van der Waals forces between layers, the material forms a layered structure; therefore, reaching a perfectly aligned edge by milling is not possible. However in simulations, a vertical hole configuration is assumed. (ii) In contrast with the experimental results, simulation results show more pronounced influences of the first and second nanocavity. This results in more contrast in the intensity and as such, the rather faint effect of the first and second hole is not as pronouncedly observed as in the experiment. (iii) The exact topology of the holes, formed by ion milling, is not cylindrical with vertical walls as considered in the simulations, but rather has some small inclined angles leading to a truncated conical topology, which has not been considered in the simulations.

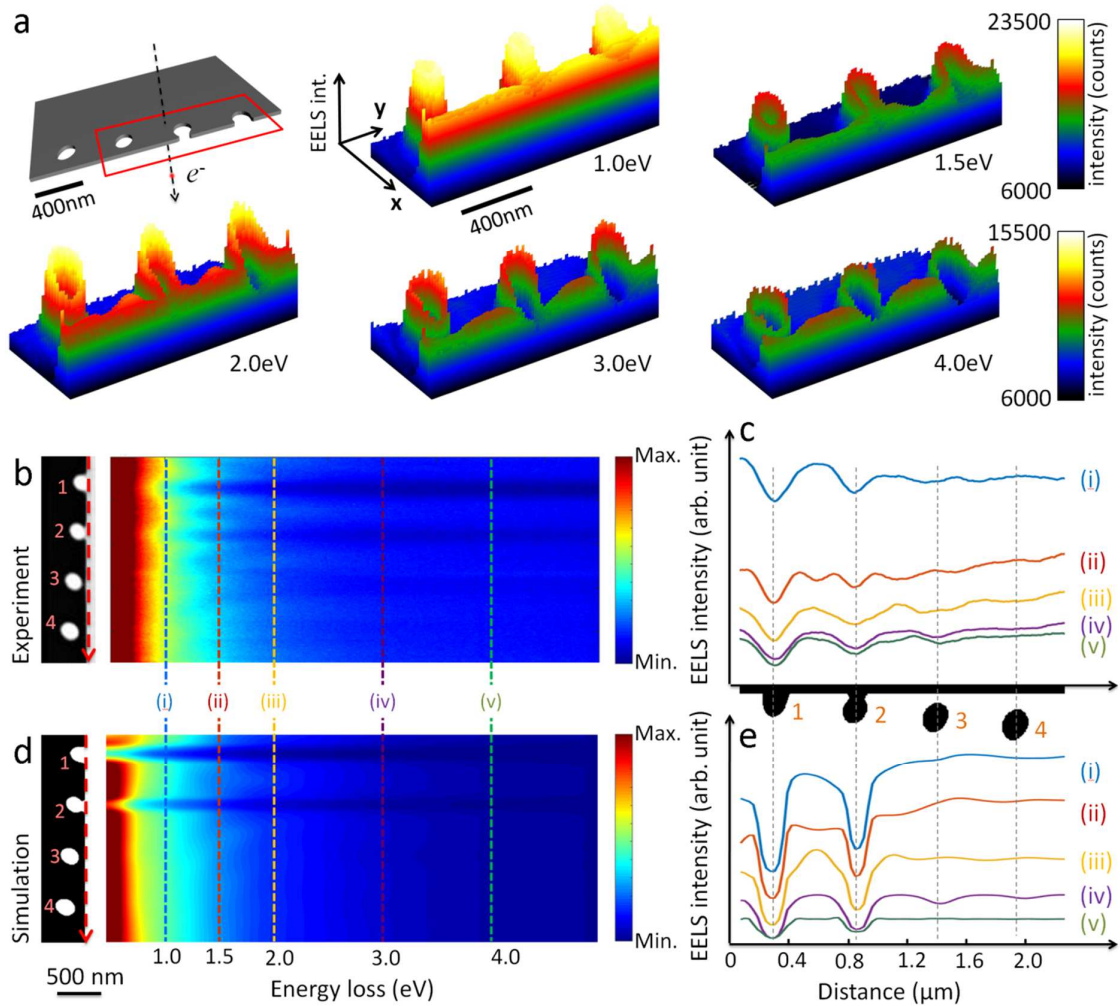

**Fig. 5 EELS measurements on the structure with holes at different distances from the edge.** (a) Schematic of the defect structure with the investigated area being marked by a red frame and slices of the 3D-data cube at the indicated energy-loss values (1 eV, 1.5 eV, 2 eV, 3 eV, and 4 eV). The color scale indicates the relative local absorption probability at the displayed energy-loss. (b, c) Experimental and (d, e) simulated EELS intensities acquired along a $Bi_2Se_3$ nanoplatelet having holes at different distances from the edge. The exact shape of the structure was extracted from the bright field TEM image and used for the simulations. Line profiles at selected energies (i) 1 eV, (ii) 1.5 eV, (iii) 2 eV (iv) 3 eV, (v) 4 eV are shown in (c) and (e).

Additionally, other reasons might also slightly alter the simulation results compared to the experiments, such as the existence of pollutant and dopants inside the material that could slightly change its optical response compared to the expected perfect $Bi_2Se_3$ single-crystalline structure.

**Discussion**

Fast electrons interact with uniaxial crystals of $Bi_2Se_3$, resulting in a variety of collective modes being excited within the bulk, at the interfaces and along the edges of the nanoplatelets. Because of electromagnetic interactions between the electrons and launched polaritons, the electrons lose energy, which is detected by an electron energy-loss spectrometer. Electrons propagating in the bulk along the *z*-axis launch longitudinally oscillating charge density waves, resulting in an absorption peak at the bulk plasmon energy of 2 eV, as well as Cherenkov radiation at energies below 1 eV, where the material is still dielectric. At the interfaces to a dielectric medium, surface modes, ranging from Channel waveguide waves[44] in the dielectric to hyperbolic polaritons[1] in the hyperbolic energy regime, transfer electromagnetic energy in lateral directions. Within the context of surface plasmon polaritons, the hybridized even and odd modes of thin metallic films are well known and extensively discussed in the literature. For an anisotropic hyperbolic material however, both S and P-polarized propagating modes are observed, with different disperions and mode profiles (Supplementary Fig. 5 and 6, and Supplementary Note 3). Although being beyond the scope of the current paper, momentum-resolved EELS measurements could unravel the dispersion diagram of surface exciton polaritons, similar to methods used to explore phonon dispersions[45,46].

In addition to surface and bulk polaritons, edge polaritons are also excited along the edges of the nanoplatelets. The dispersion of hyperbolic edge polaritons and their characteristics are intensively discussed in the Supplementary Information Note 2.

Indeed, the dynamics of edge polaritons and the phase distribution of optical near fields could convey more insight into the time-varying propagation mechanisms to understand for instance the multipole distributions of localized resonances in our cavities. In contrast to EELS experiments that show time-averaged photonic local density of states, simulations using the FDTD approach are a powerful tool to give insight into the dynamics and propagation mechanism of the polaritons and their interactions with structured defects. By exciting the specimen at certain energies with an electron at a defined impact position (marked with a white dot in Fig. 6), the propagating modes can be visualized by calculating the *z*-component of the electric and magnetic field strength in finite time intervals. When the moving electron interacts with the $Bi_2Se_3$ nanoplatelet at its edge, various energy-loss channels become possible, ranging from diffraction radiation to surface and edge polariton excitations. The dominant mechanism is the transition radiation, whereas only less than 10 % of the generated photons are converted to the propagating polaritons.

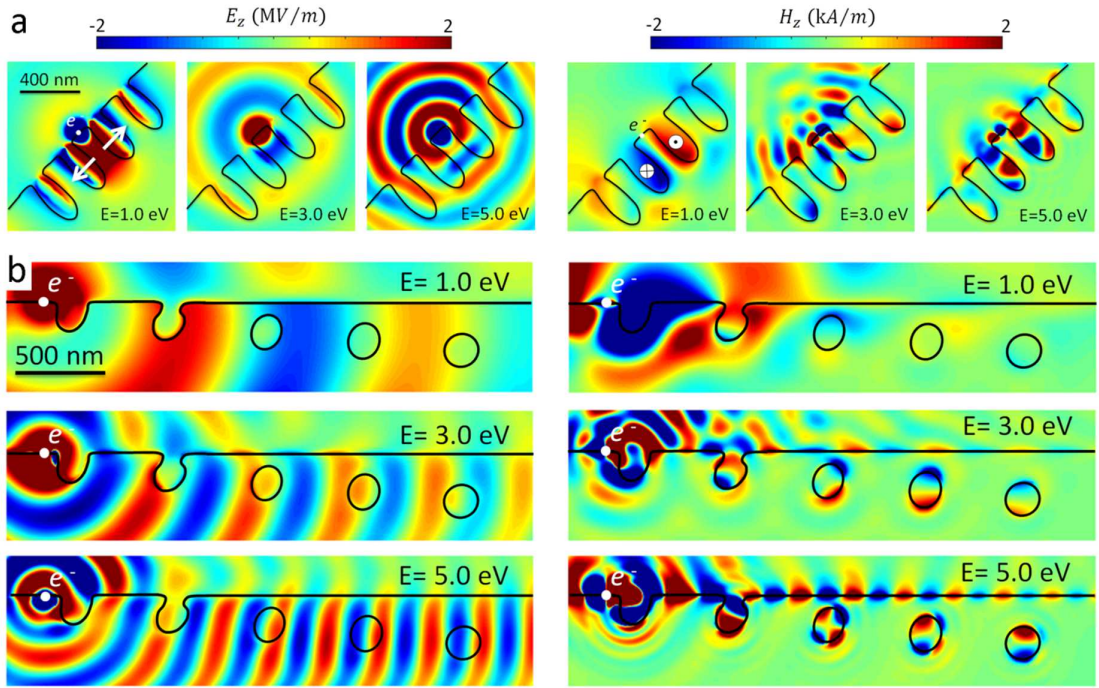

**Fig. 6 FDTD simulations for hyperbolic polaritons interacting with engineered defects.** (a, b) FDTD simulations for the investigated structures being excited by fast electrons at the positions marked in white color. The color-coded images show the *z*-component of the electric (left) and magnetic (right) field strength for different excitation energies. The images depict temporal snapshots. Movies are available online.

We start our analysis with the structure shown in Fig. 3. While propagating modes are observed along the edges of the grooves for higher energies, this is not the case at 1 eV. Instead, dipolar excitations are visible inside defects that couple to the edge polaritons. The magnetic field is confined inside the defect with asymmetric orientation related to the position of the excitation, pointing inside and outside the plane, respectively (Fig. 6 a, right panel). As such, the defect structures behave similarly to a split ring resonator at low energies. Hence the mechanism of the propagation of the polaritons at this energy is via dipole-dipole interactions. At energies higher than 3 eV, the effective wavelengths of the edge polaritons are smaller than the size of the grooves. Edge polaritons are therefore partially transmitted around the corners into the grooves, and are partially coupled to the next defect via radiation transfer. However, the latter mechanism contributes more strongly to the coupling between the defects at higher energies. Particularly at energies $E > 4.0$ eV, edge polaritons can penetrate into the grooves around bending edges as sharp as 90° and couple to the next gap as well (Fig. 6 a).

For the structure with embedded circular nanocavities at varying distances from the edge, we observe the surprising ability of the edge polaritons to cope with the presence of smaller defects. Even for the case of open nanocavities touching the edge, edge polaritons can efficiently couple to the next bridge at low-loss. At energies above 4 eV, the attenuation constant of the $HE_1$ edge mode is significantly reduced (compare Fig. 6 b with Fig. 2 b). Hyperbolic surface polaritons are purely transverse modes and do not sustain the z-component of the magnetic field ($H_z$). In contrast, edge polaritons demonstrate a strong localization along the edges (Fig. 6 b, right panels). Edge polaritons can also be localized inside isolated nanocavities in the form of dipolar and quadrupolar magnetic resonances, at the energies of 3 eV and 5 eV, respectively. In comparison with surface polaritons, the spatial distribution of edge polaritons is quite confined to the edges, which makes them suitable candidates for engineering localized nanoresonators made of $Bi_2Se_3$.

Our overall observations highlight the unique feature of optical excitations in tetradymites and in particular in $Bi_2Se_3$. Distinct from surface and edge plasmon polaritons, the visibility of the spatial interference fringes caused by the Fabry-Pérot-like resonances in the $Bi_2Se_3$ nanoplatelets are much less pronounced. As discussed here, this is due to the coexistence of various optical modes and therefore low beating frequencies caused by the phase differences between the optical modes. Moreover, exactly this competition between the different optical modes helps for a higher transmission through discontinuities, simply by facilitating the modes couplings and various energy transfer mechanisms as discussed above. Moreover, we do not observe a zigzag-like propagation of the sort normally observed in h-BN and other materials supporting phonon polaritons. In other words, the optical response of $Bi_2Se_3$ sustains a combination of plasmonic-like propagation concomitant by a large number of photonic states, supported along surfaces, edges, as well as in the bulk of the material.

Finally, we would like to point out that our observations regarding the behavior of edge polaritons support the ellipsometry model used in treating the material as a uniaxial anisotropic material, in a sense that the in-plane dispersion of the optical waves are completely isotropic. We did not observe any evidence that the behavior of edge polaritons along certain in-plane edges, regarding their orientation with respect to the in-plane crystallographic direction, is different. However, natural nanoplatelets do behave differently from long milled edges, highlighting the fact that due to the layered nature of the material, milling the structure to achieve a perpendicularly aligned edge is challenging.

**Conclusion**

Our results show that $Bi_2Se_3$ can host room temperature polaritons, similar to other classes of van der Waals materials such as transition metal dichalcogenides. The excitonic optical response of $Bi_2Se_3$, in combination with its hyperbolic behaviour, allow for the excitation of hyperbolic edge exciton polaritons[21]. The interaction between such HEEPs and engineered defects of different topology are revealed here in detail. We were able to directly visualize standing-wave patterns resulting from the reflection of hyperbolic polaritons from edges and corners, and furthermore to experimentally demonstrate steering of the edge polaritons by means of grooves and nanocavities in the vicinity of edges. Supported by numerical simulations, we could show that at higher energies, edge polaritons can avoid defects, and that unexpectedly the standing wave ratio associated with the reflection from such defects is smaller than at lower energies. Remarkably, due to the large number of hybrid optical modes, which are excited at higher energies and also the low attenuation constant of edge polaritons at $E > 4.0$ eV, embedded discontinuities are overcome by coupling between various edge and radiation modes. This characteristic is promising for the implementation of data transmission lines as interconnects between electronic counterparts, and opens up possibilities for coherent and efficient transfer of optical energy similar to topological photonics, albeit within a natural material such as $Bi_2Se_3$. Moreover, the principal possibility to tune hyperbolic polaritons in van der Waals materials via doping and defect engineering renders $Bi_2Se_3$ and related tetradymites into promising platforms for future configurable optical waveguides, both in the reststrahlen bands as discussed elsewhere[47,48] and at visible frequencies as demonstrated here. Our concept provides an excellent basis for future applications such as nanooptical circuitry, cloaking at the nanometer scale, as well as quantum technology by steering emission of quantum emitters on the nanoscale.

**Methods**

**Sample growth and transfer.** Nanoplatelets of $Bi_2Se_3$ were grown on $Si/SiO_2$ substrates using a catalyst-free vapor transport method as described elsewhere[49]. They were then transferred to a TEM grid by an all-dry approach to minimize contamination due to solvents[50]. To this end, the $Si/SiO_2$ growth substrate was gently pressed onto a holey carbon support film, while avoiding lateral movement.

**Focused ion beam sample preparation.** Structuring was performed by direct milling with a focused ion-beam. In order to achieve the required high resolution we utilized the Raith ionLine Plus system as a dedicated high-resolution ion-beam structuring tool. The holes were milled using a focused beam of double-charged gold ions ($Au^{++}$) at an acceleration voltage of 35 kV and a beam-limiting aperture with a diameter of 7 µm, resulting in a beam current of 11.3 pA. We suspect that there are little to no ion beam structuring related defects in the final structure. Firstly, the used FIB system was optimized to suppress the beam tails outside the intensity maximum of the primary ion beam, which are responsible for the parasitic milling effect. Secondly, we see little effects due to ion implantation, as we do not mill into bulk samples but into suspended individual crystal, which allows milling through the material in its entirety. The very thin holey carbon support film made structuring challenging for two distinct reasons: Firstly, larger dose than required for drilling through the $Bi_2Se_3$ material do immediately tear the support film. This holds in particular if the film exhibits holes near the $Bi_2Se_3$ nanoplatelet, which is to be structured. Milling doses therefore need to be carefully adjusted for every structuring process and were on the order of 0.05 nC/µm² for the area structuring mode. Secondly, the milling strategy, that is the dose deposition sequence, has prominent influence on the results. We suspect that this is related to heat or to stress building up in the nanoplatelet itself or at its interface with the carbon support. Circular milling strategies, in which the dose is deposited in concentric pathways, have proven to be most stable. Therefore, sharp corners and edges are difficult to fabricate in the current setup. The Supplementary Fig. 9 depicts a high resolution image of the finite grating structure, demonstrating the quality of our fabricated defects.

**ZEISS SESAM transmission electron microscope.** Experiments were performed using the Sub-Electronvolt-Sub-Angstrom Microscope[51], an instrument ideally suited for low-loss EELS investigations. It is equipped with an electrostatic Ω-type monochromator and operated at an acceleration voltage of 200 kV. The in-column MANDOLINE energy filter offers high energy dispersion at very good stability. Analytical EELS measurements are possible in the two operational modes energy-filtered transmission electron microscopy and scanning transmission electron microscopy spectrum imaging (STEM-SI), that both build up a three-dimensional data cube with two spatial dimensions and one energy-dispersive axis. For EFTEM, the microscope is operated in TEM mode, where the area of interest is illuminated with a parallel electron beam. An EFTEM image is then projected onto the screen or CCD camera by electromagnetic lenses. Images are captured at different energy-loss values by adding an offset energy to the primary electrons with a step width of 0.2 eV. This way, comparably large areas up to several micrometers can be investigated at high spatial sampling with the drawback of limited energy resolution. For STEM-SI, the microscope is operated in scanning mode with a convergent electron beam being focused to a small probe, which is scanned over the area of interest. Spatial information is achieved by correlating the z-contrast signal originating from elastically scattered electrons and the EEL-spectrum of inelastically scattered electrons to the scanning probe position. The spectrum is acquired with a dispersion of 5 meV per channel on the CCD. High-energy resolution below 100 meV is routinely achieved on a daily basis as determined from the full width at half maximum of the zero-loss peak. Independent from the acquisition mode, the color-coded intensity variation in EFTEM images corresponds to spatially resolved absorption probabilities at the displayed energy-losses, caused by the induced electric field acting back on the electron.

Thicknesses have been determined by acquiring $t/\lambda_{MFP}$ maps in TEM mode with the inelastic mean free path $\lambda_{MFP}$ being calculated with David Mitchell's mean free path estimator script (https://www.felmi-zfe.at/dm-script/dm-script-database/, accessed January 11, 2019).

Both $Bi_2Se_3$ specimen and the amorphous carbon support film are beam sensitive materials. As monochromatic electron beams were used for the experiments, the problem of beam damage was minimized through the reduced beam current. For EFTEM, a comparably large area has been illuminated with a reduced current density. However, a focused condenser can lead to a fast degeneration of the support film. In scanning mode, both specimen and the carbon film are very stable with arising hydrocarbon-contamination being the major problem.

EFTEM series have been performed using a collection semi-angle of 6 mrad with variable exposure times up to 30 s for each energy-filtered image with subsequent scaling. For one or two dimensional spectrum imaging in scanning mode, spectra at each spatial pixel position were acquired using an exposure time of 0.9 s under a collection semi-angle of 7 mrad. The spectrum was slightly spread on the CCD camera in the non-dispersive direction to improve the signal quality without oversaturating individual pixels.

**Simulations.** Binary models of the structures have been extracted from TEM images as a basis for simulation work. Here, a spatial discretization of 2 nm was used. For each of these unit cells, the anisotropic permittivity components were described by the Drude model in combination with another critical function to implement the interband transition[52]. This model can be described as

$$\varepsilon_\alpha(\lambda) = \varepsilon_{\infty,\alpha} - \frac{1}{\lambda_{p,\alpha}^2 \left(1/\lambda^2 + i/\gamma_{p,\alpha}\lambda\right)}$$
$$+ \frac{A_{f,\alpha}}{\lambda_{f,\alpha}} \left[ \frac{e^{i\varphi_{f,\alpha}}}{1/\lambda_{f,\alpha} - 1/\lambda - i/\gamma_{f,\alpha}} + \frac{e^{-i\varphi_{f,\alpha}}}{1/\lambda_{f,\alpha} + 1/\lambda + i/\gamma_{f,\alpha}} \right], \qquad \alpha = \perp, \parallel \qquad (3).$$

This function has been employed elsewhere for gold and silver[52,53]. Here we applied it to the dielectric function of $Bi_2Se_3$. The fitting parameters are given by $\varepsilon_{\infty,\parallel} = 0.082$, $\lambda_{p,\parallel} = 113.485\,\text{nm}$, $\gamma_{p,\parallel} = 88\,\text{nm}$, $A_{f,\parallel} = 9.711$, $\lambda_{f,\parallel} = 750.969\,\text{nm}$, $\gamma_{f,\parallel} = 3.138\,\mu\text{m}$, $\varphi_{f,\parallel} = -5.048°$, $\varepsilon_{\infty,\perp} = 2.485$, $\lambda_{p,\perp} = 65.81\,\text{nm}$, $\gamma_{p,\perp} = 5\,\text{nm}$, $A_{f,\perp} = 1.175$, $\lambda_{f,\perp} = 750.969\,\text{nm}$, $\gamma_{f,\perp} = 20.869\,\mu\text{m}$, and $\varphi_{f,\perp} = -4.339°$.

For the excitation by an electron probe in FDTD simulations, a charge broadening scheme was applied as described elsewhere[54,55]. In our simulations, we have combined Maxwell's equation with the Lorentz equation using the relativistic Vay scheme[56]. Therefore, the propagation of the electron is governed by its relativistic initial velocity and direction, as well as electromagnetic forces inserted on it. Nevertheless, we notice that the latter forces enforce negligible changes of the electron trajectory, as such the so-called non-recoil approximation is still valid[57,58]. The dispersion diagram of the ridge waveguide of $Bi_2Se_3$ was computed by using the finite-difference frequency domain method[59]. In order to obtain eigenvalues that correspond to physical modes, only those with phase constants less than 50 rad/μm and attenuation constants less than their phase constants were extracted.

**Author information**
Corresponding authors
Robin Lingstädt (E-Mail: r.lingstaedt@fkf.mpg.de)
Nahid Talebi (E-Mail: n.talebi@fkf.mpg.de)

The manuscript was written by RL and NT. RL has done the EELS, EFTEM and CL experiments. NT performed the theoretical and numerical investigations. The $Bi_2Se_3$ nanoplatelets have been fabricated by SM, and were structured by MH. HG and NT conceived the idea. The entire work has been supervised by NT, HG, PvA and MB. All authors have given approval to the final version of the manuscript.

**Acknowledgements**
NT acknowledges funding from the European Research Council (ERC starting Grant NanoBeam). MH and HG acknowledge financial support from the European Research Council (ERC Advanced Grant ComplexPlas), Bundesministerium für Bildung und Forschung, Deutsche Forschungsgemeinschaft (SPP1839), and Baden-Württemberg Stiftung. Work in Stuttgart Center for Electron Microscopy has received funding from the European Union's Horizon 2020 research and innovation programme under grant agreement No. 823717 – ESTEEM3. All authors acknowledge Birgit Bussmann for ultramicrotomy sample preparation, Audrey Berrier for ellipsometry measurements and Wilfried Sigle and Rainer Hillenbrand for very helpful discussions.

**Data availability**
The datasets generated during the current study are available from the corresponding author on reasonable request.

**Code availability**
The developed code used in this study is available from the corresponding author on reasonable request.